\documentclass[11pt]{article}

\usepackage[utf8]{inputenc}
\usepackage[T1]{fontenc}
\usepackage[margin=1in]{geometry}
\usepackage[round,authoryear,compress]{natbib}
\usepackage{lmodern}

\usepackage{enumitem}
\usepackage{xcolor}
\usepackage{array}
\usepackage{booktabs}
\usepackage{tabularx}
\usepackage{graphicx}
\usepackage{amssymb}
\usepackage[colorlinks=true,linkcolor=black,citecolor=blue!55!black,%
            urlcolor=blue!55!black,breaklinks=true]{hyperref}

\newcommand{\gs}[1]{\href{#1}{Google Scholar}}
\setlist[itemize]{leftmargin=1.4em,itemsep=2pt,topsep=2pt}
\setlist[enumerate]{leftmargin=1.6em,itemsep=3pt,topsep=2pt}

\title{Below the Surface: Creep, Corrosion, and Tipping Points\\
in the Social Resilience of Science\\[4pt]
}
\author{Pawel Sobkowicz\\
National Centre for Nuclear Research\\
Otwock-Swierk, Poland\\	
\url{pawelsobko@gmail.com} 
}
\date{\today}

\begin{document}
\maketitle

\begin{abstract}
By its most visible quantitative metrics: numbers of researchers, papers, journals, and institutions, science has never looked healthier, and has grown almost without interruption for three centuries. 
Yet a set of subsurface indicators points the other way: the disruptiveness of the average paper and patent is falling, progress slows in the largest fields, replication is rare and often unsuccessful.
Incentive systems can select for poor methods even when no one is actually cheating (and some people do cheat). 
Viewed from the lay society perspective, there is a curious mixture of admiration and lack of trust for scientists and their work. When asked about research as a whole, people in wealthy democracies express quite high levels of trust. 
However, there are localised ``pockets'' of distrust, behaviourally revealed (by refusal to follow the practices prescribed by science). Such cases of distrust are exploited by adaptive adversaries who manufacture doubt and weaponise it. 

We argue that a healthy-looking surface can coexist with, and mask, an accumulation of below-the-surface damage due to slow-acting, internally and externally driven pressures, weakening the social position of science. This process is best understood not as an acute shock but as a creeping crisis of the social resilience of science. And while the damage grows slowly, it may result in a sudden collapse of the research as social institution.

To make this diagnosis tractable rather than merely alarming, we import the conceptual and mathematical apparatus of two mature research fields that already study such slow degradation and sudden collapse processes: the continuum damage mechanics of creep, fatigue, and corrosion, and the ecology of resilience, regime shifts, and critical transitions. We note both the analogies and dissimilarities. From the first field we take damage accumulation, effective stress, treatment of coupled damage channels, and a networked, weakest-link failure criterion; from the second, basin geometry, hysteresis, slow-variable dynamics, and early-warning signals. We reduce the research system to four coupled,  ``health'' dimensions: exploratory capacity, verification capacity, social trust and legitimacy, available resources, and human capital. 
We identify their principal stressors, feedback loops, and add the one feature absent from both source domains: the presence of a strategic adversary that optimises the load and applies it to specific components of the complex system. 
We deliberately do not claim to resolve whether science is merely robust under strain or is approaching a tipping point; the available data are too fragmentary. Instead we argue this is the question of importance and propose what would answer it: candidate early-warning indicators (which do not yet exist and must be constructed and validated, not assumed), mechanistic agent-based models of the coupled trust–incentive–funding feedback system, adversarial red-teaming imported from epistemic security studies, and the design of potential resilience building and recovery interventions. The result is a research agenda, and an argument for why the health of science cannot be read off its own success reports.
\end{abstract}

\newpage

\tableofcontents
\bigskip

%==============================================================================
\section{The paradox of healthy-looking science}
%==============================================================================
Research as a social activity is, by its most visible metrics, flourishing. It
has been, without interruption, for three centuries. 
We are surrounded by technologies designed and built using scientific discoveries. We live longer and healthier thanks to medical advances. Our understanding of the world, from nanoscale to the Universe, is growing. We are proving mathematical conjectures from the past. 
The number of active researchers and their output grow exponentially, doubling roughly every 10--15 years, as noted by the founder of scientometrics, Derek de Solla~Price, who traced it in the accumulating volumes of the Royal Society's
\emph{Philosophical Transactions} back to the 1660s 
\citep{SollaPrice1986LittleScienceBig}. If anything, the growth seems to be accelerating. 
Pooling four bibliographic
databases across the modern era, \citet{Bornmann2021GrowthRatesModern} estimate an overall growth
rate of about 4.1\% per year (a doubling time of ${\approx}17$ years), rising to
${\approx}5\%$ since 1945; the segmented analysis of \citet{Bornmann2015GrowthRatesModern}  puts the
post-1980 rate at 8--9\% per year, and \citet{Larsen2010RateGrowthScientific}  find no sign of
deceleration over the preceding half-century. In absolute terms, STEMM researchers publish about  3.5 million journal articles annually \citep{Baker2025IsMoreScience}.
The research workforce and the number of research journals
have expanded in rough parallel \citep{Fortunato2018ScienceScience}. 
Funding
grows more slowly than the community it must support: grant success rates
fall in most countries, but despite this the general trajectory of the linked domains of scientific effort,
technological advances, economic growth, and social progress remains upward.

Two comments are due here. First, the exponential growth was long expected to end: de Solla
Price himself predicted that it would turn into a saturating
logistic curve and reach a ceiling. This prediction has, so far,
conspicuously failed to arrive \citep{Baker2025IsMoreScience}. 
Note that even such saturation would not be a sign that something is wrong: 
the production of new knowledge would continue, discoveries would be used in solving humanity's problems and challenges or just answer our curiosity. 
The persistence of exponential growth simply shows that the need for research is still not fully satisfied, 
governments and industries are willing to pour more money into research,
the production limits are not reached, and working as a scientist is attractive to an increasing number of people.

The second remark is more cautious. At least a part of the most recent scientometric
acceleration is due to changes in measurement: expanding database coverage,
the rapid rise of new producer nations, the mainstreaming of conference papers and
preprints counted as research outcome. 
Moreover there are negative  phenomena that also inflate the modern counts, for example the proliferation of paper mills. 
These ``parasitic'' phenomena distort the quantitative measures to some extent, but in our opinion, only in a limited way.

Another, this time qualitative ``measure'' of success of today's science can be gleaned from yearly reports of the funding institutions throughout the world. They are full of descriptions of successful projects and promises of more to come.
Asked about the statistics of failed projects (defined as projects that have not reached the breakthroughs promised in the corresponding grant proposals), the funding agencies have two stock answers ``none'' or ``we do not provide such information''.
I know, I tried. In a recent conversation a high ranking representative of such institution suggested thet the real answer is ``we do not know'', which seems even worse. 
While the competitive \emph{proposal} selection is frustrating, and low success rates  are  punishing (see Section~2), the
\emph{execution} success rate, namely the share of awarded grants whose final reports
declare them a success, is  close to 100\%.  Unfortunately, this near-universal
\textit{reported} success is an inevitable artefact of the aligned incentives of the involved parties, 
rather than a measure of innovation and discovery.
Once a project gets funding, the principal investigator, the host institution, and the
funding agency share an interest in recording that its milestones were met and its
promised breakthroughs delivered: the researcher earns prestige and the next
award, the institution claims credit, and the agency demonstrates that public
money was well spent. Short of gross misconduct made public, no party has an incentive to
report failure. Universities and governments alike report successes. And, of course, so do the involved researchers.
Hence the glorious picture.

Science as a community and process has been the subject of many studies, see, for example the review of \citet{Fortunato2018ScienceScience}. Our goal, while touching many aspects of its functioning, is limited. The premise of this paper is that these good health and growth metrics are a
\emph{surface} phenomenon. Exponential growth of \emph{output} is not exponential growth of
\emph{discovery}: the disruptiveness of the average paper has been falling for
decades even as their number soars (\cite{Park2023PapersPatentsAre,Wu2026IsInnovationBecoming}, see discussion below), and progress slows in the
largest fields \citep{Chu2021SlowedCanonicalProgress}. Reassuring aggregate indicators can therefore
coexist with, or even mask, an accumulation of \textit{subsurface damage}. That
divergence: healthy surface, degrading substructure, is the signature of the
phenomena we analyse, and the reason to look at the situation through a crisis
lens. We are going to ask \textit{whether science as a social activity can withstand}, not just acute shocks, such as global natural catastrophes, wars etc., but also slowly changing societal factors, that determine the environment in which research is practiced.

\textbf{We want to look at the \textit{resilience} of research as a complex social system}, understood as the
capacity to absorb disturbance and still retain its essential function and identity
\citep{Holling1973ResilienceStabilityEcological,Folke2006ResilienceEmergencePerspective,Carpenter2001MetaphorMeasurementResilience}. It is not limited to mere stability, but the ability to persist
through change and, when necessary, to reorganise. And our focus is not on acute disturbances (wars, pandemics, economic crises) but on the slow-changing social circumstances that influence how research operates in reality.

The plan of the paper is as follows: we first discuss the already recognised signals of problems hidden behind the facade of exponential growth (Sections~2-4). Section~5 discusses the aspects of the potential failure of science as a social system in terms of a connected network of characteristic components and functionalities and their failure modes. Section~6 briefly introduces the two research domains, which extensively deal with resilience of complex systems: mechanical engineering and ecology, together with the basic ideas we shall be borrowing. Sections~7--8 present the basic framework of the  models based on these two approaches, as well as the proposed research agenda.

%==============================================================================
\section{Subsurface degradation I: the internal creep}
%==============================================================================
The exponential growth of science may be a sign of its health, but it may also indicate that something is wrong, out of control. That quality does not follow the explosive growth of quantity. 
A gap between exponentially growing \emph{output} and
stagnant \emph{discovery} rate would be such a signal. 
The disruptiveness of the average
paper and patent (their tendency to push a field in genuinely new directions,
measured by the CD index) has fallen by on the order of 90\% for papers
(1945--2010) and 80\% for patents (1980--2010) \citep{Park2023PapersPatentsAre}. The reported values depend on the scientific discipline.  This claim (in particular, the large value of the decline) is
contested on grounds of the validity of the CD index definition and use \citep{Kim2026UncoveringSimultaneousBreakthroughs,Leibel2023WhatDoWe,Bentley2023IsDisruptionDecreasing,Petersen2024DisruptionIndexIs,Macher2024IsThereSecular}, in response \citet{Wu2026IsInnovationBecoming} published a literature review (not peer reviewed at the moment of writing), listing analyses corroborating the decline in disruptiveness of research publications. 
Additional supportive argument is provided  by \emph{slowed canonical progress in large fields} \citep{Chu2021SlowedCanonicalProgress}, 
and by a dramatic decline in research productivity, for example, aggregate US research productivity declined by a factor of 41 since the 1930s, about
5\% lost per year \citep{Bloom2020AreIdeasGettinga}. The trend has a demographic signature too:
analysing 65 million papers, patents and software products, \citet{Wu2019LargeTeamsDevelop} show
that small teams \emph{disrupt} while large teams \emph{develop} existing ideas, so
the universal drift toward ever-larger teams creates a drift from breakthrough
toward incremental results. 

Why should the disruptive share (breakthrough and high-impact discoveries) fall even as publications soar? 
There are two potential major explanations. First, that ``there is less left to find'', the construction of the body of knowledge is almost finished. Fewer revolutions are expected because the big things are discovered. Almost all the growing effort and output is focused on filling the gaps \citep{Horgan1996EndScience}. This, of course, recalls the predictions by Albert Michelson and Lord Kelvin, about the end of physics, stated just before the greatest revolutionary period at the beginning of 20th century.  Yes, as we mentioned in the opening section, it is possible that certain research fields  are truly ``mature'', so there are no revolutionary discoveries to be made,  and all that is left to be done there is more precise measurements. 
But this certainly is not the case for all disciplines. Recent years brought fundamental, paradigm changing advances in such diverse fields as biology, astronomy, quantum computing, mathematics, artificial intelligence\ldots  
Yet these discoveries are overshadowed by the mountainous pile-up of run-of-the-mill papers.

Several accounts point to a different reason for this ``law of diminishing returns'': the cost and effort of reaching the frontier, where breakthrough discoveries happen, is rising. There is the burden of knowledge barrier, identified by \citet{Jones2009BurdenKnowledgedeath}: as knowledge accumulates, each innovator must learn more before reaching the frontier, so they have to specialise, start the truly innovative work later, and cluster into teams, which lowers individual innovative capacity. 
\citet{Chu2021SlowedCanonicalProgress} describe the ``ossification mechanism'': past a certain field size, progress slows not from lack of things to disrupt, but because to reach the frontier one has to get through vast areas of accumulated knowledge. There are too many papers per year, so citations flow back to entrenched canonical works and genuinely new ideas sometimes cannot get attention. 
Similar ``natural'' slow-down arguments were raised even before Horgan's doom predictions. \citet{Rescher1978ScientificProgress} argued that science becomes more expensive, each increment of fundamental advance requiring exponentially more resources, but not due to the lack of things to be discovered. His predictions were validated in recent studies.

But there is another part of the answer to the question why we have relatively less high impact discoveries. It is more troublesome, because it points at a negative selection internal process:
the systems allocating publication and funding actively select \emph{against}
novelty, creating a \textit{mediocrity bias}. In a field experiment, grant evaluators systematically
rated the most novel proposals \emph{lower} \citep{Boudreau2016LookingLookingKnowledgea}.
There is a real risk of being too ``revolutionary'' to be understood by the reviewers.  A rational
response to such selection is to ``conform and be funded'' \citep{Nicholson2012ResearchGrantsConform}. 
This creates a feedback loop strengthening the bias.
Formal and
simulation models confirm and develop these observations: under plausible assumptions competitive peer
review tends toward the ``selection of the average'' \citep{Thurner2011PeerReviewWorld}.

Grant peer review, now a dominant way of securing funds for research,  was presented as a tool for slowing down scientific progress (so that ethical progress can catch up),  by Leo Szilard almost 80 years ago in his satirical (but quite serious) story \citep{Szilard1948MarkGableFoundation}. Five decades later, \citet{Horrobin1996PeerReviewGrant} 
called the system ``a harbinger of mediocrity''. 
Yet it has become, in many countries, the main source of research funding, and is presented to the general public as the only fair and objective method of funding the ``best scientists''. 
An agent-based model of competitive,
peer-review-based funding exhibits the mechanism directly: unless reviewers are
unusually tolerant of proposals unlike their own, the scoring process drives a
``regression towards mediocrity,'' suppressing innovation while entrenching
self-reinforcing cliques \citep{Sobkowicz2015InnovationSuppressionClique}. 
The low disruptiveness drift is not just simply a result of the growth of the sheer volume of research, but is strengthened by the selection filter.

There are other signals that indicate that something is wrong. In psychology, a major   
replication effort
reproduced only 36 out of 97 published psychology results by the significance criterion  \citep{OSC2015EstimatingReproducibilityPsychological}. 
This is consistent with the structural argument
that most published findings may be false under prevailing designs
\citep{Ioannidis2005WhyMostPublished}. 
The assumed ``self-correcting'' nature of science, based on multiple researchers checking up and correcting others has been almost forgotten in practice. The funding agencies push for \textit{novelty}.
In social sciences, where the contexts significantly vary and replications are absolutely necessary, they are, in fact, extremely rare \citep{Makel2012ReplicationsPsychologyResearch,Makel2014FactsAreMore}. 
The rate is higher in medicine, but in most studies it remains low.
This ``abandonment'' of the fundamental mechanism of research through biased incentive system has prompted calls for introducing special funding devoted to replications \citep{Koole2012RewardingReplicationsSure,Derksen2024ReplicationStudiesNetherlands}. 
We should remember that replicability is a necessary procedural requirement for reproducibility: actually confirming the reported findings.

The negative  dynamics are self-reinforcing: an evolutionary
model shows publication-based selection propagating poor methods \textit{with no conscious
cheating} leads to a ``natural selection of bad science'' \citep{Smaldino2016NaturalSelectionBad}.
All these effects are
sustained by the hypercompetitive climate \citep{Edwards2017AcademicResearch21st}.
These pressures are transmitted through how researchers and institutions are now
evaluated. Careers and rankings turn on scientometric parametrisation, publication
counts, journal impact factors, the $h$-index, and citation tallies,
directly driving hiring, promotion, and funding decisions  \citep{Hicks2015BibliometricsLeidenManifesto,Mueller2017ThinkingIndicatorsExploring}, under a \emph{publish-or-perish} regime that rewards volume and conformity over
depth \citep{Edwards2017AcademicResearch21st}. 
The result is a \emph{Goodhart dynamic}: ``when a measure
becomes a target, it ceases to be a good measure''  \citep{Strathern1997improvingRatingsAudit}. 
Once a proxy for quality is optimised
 as an end in itself, the correlation that made it
useful decouples and effort migrates to the metric rather than to the underlying
good: salami-slicing, least-publishable-units, citation trading, and strategic
novelty-avoidance being the visible symptoms, with academic publishing metrics now
a textbook case of the law in action \citep{Fire2019OverOptimizationAcademic}. 
The healthy-looking growth
of Section~1 is thus partly an artefact of measurement pressure: it is the very
quantity the incentive system optimises, which is precisely why it can keep rising
while individual disruptiveness, reproducibility, and novelty fall.

The quest for universal and objective measures of discovered effects \textit{importance} and \textit{validity} is a laudable one, and reliance of strict statistical methods is natural. However, it has led to over-reliance on a specific measure (the $p$ value), and resulted in a negative practice of \textit{p-hacking} (another example of Goodhart dynamics). \citet{Simmons2011FalsePositivePsychologyb} found that undisclosed researcher flexibility can inflate the nominal 5\% false-positive ($p=0.05$) rate to as high as ${\sim}60\%$. The phenomenon is widely present in psychology \citep{John2012MeasuringPrevalenceQuestionable}, but may be found in other disciplines \citep{Head2015ExtentConsequencesP} and in observational studies \citep{Bruns2016PCurvePhacking}. 
It is a concrete example of how bad methods propagate under forced selection \citep{Smaldino2016NaturalSelectionBad}. It distorts the integrity of the accumulated knowledge base  by flooding the literature with significant-looking noise: meaningless results look like findings. Moreover, p-hacking does not indicate bad intentions on the part of the researcher(s). It is a consequence of methodology requirements imposed by journal editors. But  data-dependent analysis choices can distort inference even when no such intentional fishing is present and a hypothesis is properly fixed in advance \citep{Gelman2013GardenForkingPaths}. 

The ``universality of (awarded) grant success'' mentioned in Section~1 is much harder to document, because, as we noted, it's not in anyone's interest to report failure. Yet, common sense suggests that truly novel research is highly risky, and should carry a high ratio of failures (defined as uncorroborated hypotheses, failed experiments, lack of progress, etc.)  These ``failures'' are essential for the progress (which ties them with the need for reproducibility).  
Measuring the effect is very difficult, because the evidence is circumstantial rather than direct.
Principal investigators \emph{over-forecast} the scientific and
operational outcomes of their own projects \citep{Benjamin2022PrincipalInvestigatorsOver}; 
scientific impact
is a \emph{decelerating} function of funding, so additional money does not
predictably yield additional discovery \citep{Fortin2013BigScienceVs}; and the whole apparatus
of ex-post monitoring exhibits the pathology of an \emph{audit society}, in which
rituals of verification certify financial and procedural compliance while the
substantive question, was the promised scientific breakthrough achieved, goes unasked, and is
arguably unanswerable by the bureaucratic machinery that poses it
\citep{Power1999AuditSociety,Strathern1997improvingRatingsAudit}. The pattern scales: the deliberate over-promising
that secures approval and then reports success is the research analogue of the
``strategic misrepresentation'' documented in large public projects
\citep{Flyvbjerg2007CurbingOptimismBias}, and it operates at programme level too: the European
Union's Lisbon Strategy set a 3\%-of-GDP research-investment target and the goal of
becoming the world's most competitive knowledge economy by 2010, both quietly
missed even as interim reporting stayed upbeat. What is absent is the base rate any
commercial venture portfolio takes for granted: that most exploratory bets fail
and a few that succeed pay for the rest. A research enterprise devoted by definition to the
risky pursuit of the unknown, yet whose every funded grant succeeds, is not
reporting on that search; it is reporting and being audited on the reliability of its own paperwork.

One particular aspect of the intense internal competitive pressures is the ``publish-or-perish''
phenomenon \citep{Grimes2017ModelingScienceTrustworthiness,Parchomovsky2000PublishPerish,Garfield1996WhatIsPrimordial,Siegel2010BattlingPaperGlut} and its consequences. In addition to creation of flood of papers, necessary for researchers' survival and promotion, it encourages tactics decreasing quality of publications, substituting quality by quantity, such as the ``salami slice'' way of multiplying number of publications or even increasing  bias against publishing negative results  \citep{Fanelli2010DoPressuresPublish,Fanelli2012NegativeResultsAre}. We note, however, that in the later analysis, \citet{Fanelli2016Researchers’IndividualPublicationa} claim that ``\textit{the widespread belief that pressures to publish are causing the scientific literature to be flooded with salami-sliced, trivial, incomplete, duplicated, plagiarized and false results is likely to be incorrect or at least exaggerated}''. From our point of view (of creeping pressures) the difference between the two conclusions is fundamental. We argue that even small, but persistent processes can lead to large scale deterioration of quality.

But the publish-or-perish stress is not purely internal to scientists and their research institutions. 
It is amplified by publishers, who found ways to effectively monetise it, by creating new journals. 
They even abused the public and governmental push for  Open Access publishing, aimed at providing universal and free access to research results. In response, the publishers created new business models. 
Afraid of losing subscription revenues, they introduced Article Processing Charges (APC) for ``gold OA'', initially quite small, but now increasing astronomically. These charges are coupled with  extended embargoes for the costless combinations of ``green OA'' and subscription-based journals, to guarantee the profits.  Moreover, the most time and effort consuming work related to publication of research materials, namely peer review, is pushed by publishers onto the research community and performed without remuneration. Thus, publish-or-perish culture is truly a goldmine for the publishers.

The business is so good and self-replicating, that  the number of predatory journals is growing explosively \citep{Beall2012PredatoryPublishersAre,Shen2015predatoryOpenAccess}.
Their presence, characterised by almost no critical peer review leads to lowering the aggregate quality of publications: external activity riding on internal pressures and driving the drop in average quality.
While this external response to internal career competitive requirements probably should not be called an attack on science (Section~4), it does involve intentional manipulation of the situation. The publisher isn't an adversary in strict sense, it isn't trying to damage science (because it is a source of its business), but rather it is a rational profit-maximiser, whose incentives happen to be misaligned. Yet, the damage they cause is real. 

There are efforts to respond to this abuse (diamond/platinum OA models, mandatory shortening of the embargo periods), even governmental pressures on the major publishers. But these efforts are faced with significant problems \citep{Grudniewicz2019PredatoryJournalsNo}, and no solution is in sight (yet).

%==============================================================================
\subsection{Funding and effects: between internal and external influences} \label{sec:funding}
%==============================================================================
We have already noted the increasing competitive/financial pressure that researchers feel everyday.
While the number of researchers and the number of published papers grow exponentially, they are in close correlation. Taking into account that the average number of collaborators has also increased, and adjusting the counts for co-authorship, the publication rate of scientists in all disciplines has not increased overall, and has actually mostly declined \citep{Fanelli2016Researchers’IndividualPublicationa}. What is worrying is the decline of high-impact, breakthrough papers, already noted. Research effort is rising substantially while research productivity is declining sharply. Total output \textit{per capita }is relatively flat, but the corresponding \textit{per capita} rate for breakthrough discoveries actually falls. A widely used example of Moore's Law's famous doubling is cited in this context \citep{Bloom2020AreIdeasGettingb}: the number of researchers required today to achieve doubling of processor power is more than 18 times larger than in the early 1970s. We note, however, two aspects of this particular example of a slowdown: the exponentially rising costs of chip production technologies as the metric scale of the devices decreases, and the fact that we are now nearing atomic-scale limits, where quantum tunnelling and lattice spacing impose hard physical floors. So Moore's Law is a case where both explanations (actual physical limit and growing unit costs) operate simultaneously and are hard to disentangle. 

The third quantity, rising in absolute terms, is funding. At the largest scales (comprising of all disciplines) funding drives headcount. Researcher numbers track R\&D expenditure through the labour market \citep{Stephan2015HowEconomicsShapes}. Which again means that the three measures: funding, people, papers follow similar growth paths. But when we look  at the high impact/breakthrough progress, the situation changes. \citet{Fortin2013BigScienceVs} observe that returns to funding are diminishing. At the individual level, impact is positively but only weakly related to funding, and is generally a decelerating function of it, so impact per dollar is lower for large grant-holders. 

While, in general, people follow the money, and produce roughly similar number of papers, there are exceptions. In some research fields the number of researchers grows much faster than the funding, leading to hypercompetition (for biomedical research this has been documented by \citet{Alberts2014RescuingUsBiomedicala}).

Of course, the salary costs (the key source of the proportionality) are not the only ones. In some disciplines, the cost of the infrastructure necessary to perform experiments is a dominating factor. Two key examples are high energy physics (where the total cost of the Large Hadron Collider is estimated at over 20 billion dollars, with construction phase costs of 4.3 billion Swiss francs, and roughly 1 billion dollars per year in operational costs); and fusion as a source of clean energy (ITER creation costs are between 18 and 22 billion dollars). Yet no progress would be possible without these investments. The cost in other disciplines is typically smaller, although can be very high (for example the total cost of bringing a new oncology drug or antibiotic to market is of the order of a billion dollars). Sometimes, technological progress may also result in a drastic decrease of costs. Such was the case of genome sequencing. The original Human Genome Project cost was about 2.7 billion dollars, while today, research-grade sequencing cost has dropped to less than 500 dollars.

The fact that many governments are willing to provide exponentially growing sums of money to fund research activities (Poland is an exception, with funding that stalled in the past decade) indicates that they believe there is a value in pursuing them. Whether the reason is technology development, which is hoped to recover the funds through economic growth, new drugs that help people live healthier and longer, new military techniques and weapons, or even keeping a part of the population (namely: researchers) quiet and not causing trouble. In democratic societies, such governmental support means that the growth of science funding is accepted by the voters, or at least that they do not care. 
Which brings us to the second domain of the subsurface degradation of research as a process: pressures and damage that originate in the world outside the scientific community, including trust in science, belief in its relevance and importance, and actual use of its results.

%==============================================================================
\section{Subsurface degradation II: the external corrosion}
%==============================================================================
Generally speaking, despite the turbulences due to the pandemic, the trust is science remains relatively high. the best current
evidence supports  this: a preregistered 68-country survey of almost $72000$ respondents
finds trust in scientists \emph{moderately high} and narratives of widespread
distrust largely unsupported \citep{Cologna2025TrustScientistsTheir}. 

Similarly,   \citet{2026TrustScience}, which collects multiple European perspectives on the issue and opens with mentioning danger signals on threats to science, yet states ``\textit{no immediate crisis of trust in science can be identified [\ldots] Eight out of ten Europeans view the influence of science and technology as positive, expecting innovations to benefit everyone and serve the public good.''}.

So, is the situation good, are people trusting research and researchers, or are there any reasons for worry?

First, the issue of public trust is far from compressible to a single numerical measure. 
\citet{Iordanou2026PublicTrustScience} provides an extensive literature review of the topic, including changes in public perception after Covid-19 pandemic, listing multiple aspects of the processes invloved, from individual to social. 

The threat is not due to general loss of trust in science  but its \emph{localised corrosion}.
Tellingly, it is often \emph{deepest in the wealthy democracies} that produce the most science.
Moreover, while support for science in general is often a socially expected verbalised attitude, these patches of mistrust and rejection mostly involve science-rejecting practices that have concrete impacts on individual and societal life.

Such distrust in science (or in specific areas of science) is often explained through the \textit{knowledge deficit model}, which is a claim about cause: that distrust/false belief stems from an information gap, remediable by supplying facts. 
\citet{Wynne1992MisunderstoodMisunderstandingSociala} argues that people are not simply lacking data; they are selecting which data to believe, making relational and value-laden judgements about whose knowledge to credit.

Consider four examples of such discrepancy between general approval claims and practical decisions. \emph{Vaccines}: globally only about 7\% disagree that vaccines are safe
(about 79\% agree they are safe), but the sceptical share rises sharply in rich democracies, 22\% in
Western Europe and fully one-third (33\%) in France, one of the highest 
worldwide \citep{2019WellcomeGlobalMonitor.}. 
It is important to note that \textit{antivaxxers} do not see themselves as anti-scientific. 
In fact, much antivaccination propaganda is presented as supporting science \citep{Kata2010PostmodernPandora’sBox,Kata2012AntiVaccineActivistsa,Moran2016WhatMakesAnti}, while pro-vaccine arguments are positioned by antivaccine propaganda as serving corporate greed or governmental control goals, rather than as scientifically valid  \citep{Bricker2018PostmodernMedicalParadigm}. 
In many cases, the anti-vaccination activists are very well versed in (selected!) literature and arguments, while a large part of the vaccinated population has only superficial knowledge of how vaccines work. So vaccine hesitancy is not a case of knowledge deficit.
In other words, it is antivaxxers who see themselves as ``true to science principles'', which can potentially impact other scientific fields. 

\emph{Climate}: despite a scientific consensus
near 97\%, only about half of Americans (${\approx}48\%$) attribute warming mostly
to human activity, a figure essentially flat since 2016 \citep{Kennedy2026AmericansAreincreasinglyPessimistic},
while an estimated 15\% deny outright that the climate is changing \citep{Gounaridis2024SocialAnatomyClimate}, along a sharp partisan gradient (note that anti-vaccination does not show a single-party bias, although the reasons for hesitancy are different for liberals and conservatives).  

\emph{Perceived
corruption}: a majority of Americans (58\%) say industry funding makes them trust
research findings \emph{less}, and no more than 19\% believe scientists are
transparent about conflicts of interest with industry \citep{Funk2019TrustMistrustInamericans’}. 
In the EU, roughly half of the public agrees that scientists cannot be trusted on controversial issues because they depend on industry money \cite{Research2025EuropeanCitizensKnowledge}.
The ``science is bought'' intuition is thus a majority position, not a fringe one.

\emph{Pseudoscience}: roughly two-fifths of Americans grant astrology scientific
credence (about 40\% call it ``sort of'' or ``very'' scientific), with the
youngest cohorts the most credulous \citep{NSB2020ScienceTechnologyPublic}. 
In Europe 34\% considered homeopathy as scientific, 41\% astrology (which might be due to confusion with astronomy, as the question with ``astrology'' substituted by ``horoscopes'' gave \textit{only} 13\% of answers considering it scientific) \citep{Eurobarometer2005EuropeansScience}. 
We note here that these are not literacy-deficit-based measurements. They are effects of \textit{rejection} of scientific/rational argumentation, made worse by media ``responding to the needs of the part of population'' and giving the topics legitimacy and reach, ``normalising'' them. 
The drop in expressed belief from astrology to horoscopes partially confirms this view: (some, fortunately not all) newspapers carry horoscopes sections, but they do not call it astrology.

Crucially, these concrete examples of anti-scientific reasoning (and subsequent decisions pertaining to health, behaviours, political choices) coexist with a high \emph{average} of expressed trust. 
This is our point: reassuring statistics obscure a disturbing practical weakening of the absorbance of science in social life.

The erosion of real trust is not only something done \emph{to} science from outside (why we used the word corrosion). Some of it
is self-inflicted. In the intensifying competition  for funding, prestige, and
recognition, researchers and, especially, their institutions routinely overstate
the importance and reach of their findings, most visibly in press releases: a
study of 462 university biomedical releases found that a third or more exaggerated
the causal claims, health advice, or human relevance of the underlying paper, and
that these exaggerations passed largely intact into the resulting news \citep{Sumner2014AssociationExaggerationHealth}, 
while the promotional vocabulary of the literature itself
(``novel,'' or ``unprecedented,'' or  ``robust'') has risen markedly over recent decades
\citep{Vinkers2015UsePositiveNegativea}. 
Announcements of imminent revolutions: new materials that will
transform engineering, cures just around the corner, are made lightly and quickly
forgotten, but the cumulative effect of repeated, unfulfilled promises is
corrosive: hype spends down public trust, and its overload leaves a residue of
disappointment that attaches to science as a whole \citep{Master2013HypePublicTrust}.

A newer pressure is only beginning to be charted: generative AI. If a plausible
answer to any question is a prompt away, the perceived need for slow, credentialed
expertise may decline: ``why consult researchers when one can ask ChatGPT?'' The
concern is not merely hypothetical. Use of generative AI as a source of
science-related information roughly doubled between 2023 and 2024, and a two-study
survey found that people with less understanding of how AI
systems work, were markedly more likely to revise their
decisions to match the AI's output, often \emph{over} the stated views of experts
or reliable sources \citep{KleinAvraham2026WhenIgnoranceInduces}. Fluent, authoritative-sounding, yet
sometimes fabricated answers can thus simultaneously bypass scientific
institutions and reinforce misinformation \citep{Joseph2025GenerativeIllusionHow}. The size and even
the sign of the net effect on trust remain contested and under active study, but
the mechanism, displacing expert judgement with frictionless machine answers, is
a plausible new channel of hollowing that any framework here must accommodate.
This issue will become more important as AI systems get better at filtering out false or unverified information and hallucinations, a goal actively pursued by all major service providers. When the quality of AI-provided information surpasses that of web searches (or even human curated data sources), while still being ``tuned'' to individual capacities and preferences of the people who ask for the information, then it would become the ultimate source in the eyes of the public, making other sources (including science and researchers) seem obsolete.

Two facts are important to note. First, much of the loss of trust is discipline/topic specific, not
uniform decline, and due to social polarisation: the politicisation of specific 
findings splits the public along
partisan lines \citep{Gauchat2012PoliticizationSciencePublic}, in an attention-scarce, 
fractionated information environment where truth-orientation competes with attention-capture
\citep{Lewandowsky2017MisinformationUnderstandingCoping}. 
Second, we deliberately avoid the \emph{deficit
model}, the assumption that distrust is mere public ignorance of ``how science
works.'' Trust is relational and co-produced; treating it as a knowledge deficit
is both empirically weak and normatively misleading \citep{Wynne1992MisunderstoodMisunderstandingSociala}.

%==============================================================================
\section{Direct attacks on science}
%==============================================================================

In the analyses presented above, we have deliberately used vocabulary borrowed from engineering and 
materials science: creep, erosion, corrosion, etc. However, unlike simple degradation or fatigue in 
metals, part of the stress or load on science is applied \emph{intentionally}. 
Before we dig into the matter, we exclude from the considerations any physical or cybernetic acts of aggression on specific research-related targets: scientists, universities or funding bodies. 
By ``attacks'' we mean only intentional \textit{social} activities, directed against such institutions, often remaining within legal limits, but nevertheless creating obstacles to research.

For example, some research topics are likely to become targets of mis- and dis-information campaigns (e.g in the form of anti-vaccination movements, climate change denials, or alternative cancer treatment proponents). 
These activities may involve in addition personal attacks and accusations of misconduct, enhanced scrutiny, and time-consuming ``freedom of information'' requests. It is often impossible to draw a hard boundary between legitimate concerns and malicious ones. 
Specific actors (of various provenance and self-interests) focus their activities on 
specific  research domains, topics, arguments, exposing the weakest elements of science as social activity \citep{Barloesius2025WissenschaftsreflexionWhatIs}. Some attacks come not from members of society at large, but from the politicians and governments, for recent US-centered discussions see  \citet{Bosman2025ResilienceOpenScience,Briscoe2025ScienceSiegeProtecting}.
The mechanical analogy still holds, but with an acknowledgment that the stress, or corrosive influences are not applied uniformly to the whole structure. Parts of the construction may be overexposed to defect-inducing conditions (e.g. locally higher concentration of corrosive substances, or vibration localised in a small part of the construction). We note that in the mechanical engineering analogy, we are not asking the question whether the local deterioration is  a result of conscious action of someone, or just an unplanned condition.

An example of such exposed and abused part of the process of attacking science as a social process is described below.

Honest doubts, need for verification, conflicting (partial) results, competing theories are necessary components of scientific practice. Without them it would not be science. 
But the same elements may be misrepresented and misused as a strategy \citep{Oreskes2010MerchantsDoubtHow,Oreskes2017ResponseOreskesbeyond}. 
Conspiracy narratives about science are
weaponized for political mobilization \citep{Bergmann2024WeaponizingConspiracyTheories}; and state and
automated actors amplify both sides of scientific controversies to corrode
consensus \citep{Broniatowski2018WeaponizedHealthCommunicationa}, exploiting the same diffusion dynamics that
make falsehood travel faster than truth \citep{Vosoughi2018SpreadTrueFalse}. This is why we frame
the problem partly in the language of \emph{epistemic security}: a society's
capacity to protect the processes by which reliable knowledge is produced,
distributed, and assessed, against both external threats and internal
vulnerabilities \citep{Seger2020TacklingThreatsInformed}. The presence of an active and adaptive adversary is the
single most important difference from the physical and ecological cases, and it
requires a distinct approach (threat modelling, red-teaming).

Some adversaries may also work with more general aims: the erosion of
expertise as a recognised element of social structure, thus disturbing the reliability of information/decision flows. 
\citet{Nichols2017DeathExpertiseCampaign} diagnoses a ``death of
expertise'', as an intellectual egalitarianism in which lay opinion and specialist
knowledge are treated as equally valid, fuelled by easy information access
and a customer-service model of higher education. Empirically,
resistance to expert consensus is strongly predicted by anti-intellectual
sentiment \citep{Merkley2020AntiIntellectualismPopulism}, and it is increasingly \emph{politically sorted}:
through cultural cognition, people credit as a genuine ``expert'' only those whose
conclusions align with their group's values, so that even perceptions of what
scientists believe split along partisan lines \citep{Kahan2011CulturalCognitionScientific}. Combined with
science-related populism, the claim that ordinary people's common sense should
override an out-of-touch academic elite \citep{Mede2020ScienceRelatedPopulism}, this converts factual
disputes into identity contests, in which the adversary need only supply
credentialed-looking counter-experts for the ``right'' side. Expert status is then
granted or withheld as a marker of political belonging rather than of knowledge.

%==============================================================================
\section{From symptoms to a problem: is creeping disruption of science real and dangerous?}
%==============================================================================
Sections~2--4 catalogue several negative characteristics of modern day research. 
These symptoms do not, by themselves, establish a crisis.
Two ways of looking at them are possible, and distinguishing them is the central question of this
paper. In the first view, research is \emph{robust}: it has always been competitive,
uncertain, and politically contested; peer review, replication, and open debate
are self-correcting mechanisms that have absorbed worse misuse; hypercompetition and
public skepticism are chronic conditions, not portents of failure.  
Similarly, the exaggeration of the claims of scientists and their institutions, 
and announcements of successes of the funded research projects may be treated as examples of universal human tendencies, present in every social activity, and not something that one should worry about. 

The second viewpoint, already briefly mentioned, is that the apparent healthy growth and reported successes of research as social endeavour are misleading.     
That the processes described above, while not catastrophic separately, can accumulate and combine, making the core foundations of science weaker, and leading to a collapse.

One can recapitulate the list of the growing signals (or sources of danger):
relative decrease of funding (compared to number of researchers/institutions/projects); increasingly precariat status of researchers (especially younger ones); lack of replication/verification studies due to focus on ``novelty''; publication deluge making it very difficult to monitor, follow and find true breakthrough results, false ones, or AI-generated ones \citep{Bernard2026AiGeneratedScientific}; growing social distrust and doubts about researchers' motivations, ethics of the validity of their results.  All of these factors can weaken the position of science in social life and its ability to solve important social problems. 

\textbf{The question is: how dangerous are these symptoms?} Is research (its practitioners and institutions)  
robust enough to weather shocks, like pandemics, wars, terrorism, environmental catastrophes or economic disasters?
Such robustness of research institutions and processes to catastrophic events is, in itself, an important subject of research. 
The war in Ukraine has shown that even in extreme circumstances research can continue, although its focal points are shifted. 
here, we are more interested in another question: \textbf{is science resilient enough to survive the slow, continued erosion of the foundations, and continue without serious disruption?} 
It is not whether science faces problems, which is universally recognised, but
whether these coupled--partly systemic, partly intentional factors--are eroding the
position of research as a social institution.

In the following discussion we shall use analogies from two research domains that have already dealt extensively with such slow-changing, apparently innocuous influences that can have catastrophic consequences: mechanical engineering and ecology. Next Section shall be devoted to the introduction of these analogies.

\textbf{Our working hypothesis is that the observed negative symptoms of current research processes described above are signals that the whole system is 
under a specific threat, \emph{creeping disruption}: a slow, distributed process, foreshadowed by precursor events, recognised only partially and insufficiently addressed, the defining profile of a
\emph{creeping crisis}} \citep{Boin2020HidingPlainSight,Boin2021UnderstandingCreepingCrisis}.
(The focus on the slow, subsurface processes does not exclude the potential presence of abrupt and acute forms of crisis, such as war, economic downturn, social unrest, or pandemics, which we recognise but do not consider here.) However we would like to include the adaptive adversaries (mentioned in Section~4), who often  deliberately accelerate the disruptive processes mentioned above, for example by sowing doubts to specific social populations.

One note to avoid confusion: the ``creeping disruption'' term 
for the gradual erosion of the research system's social foundations has to be
distinguished from the desirable  \emph{intellectual disruptiveness}, creation of novel ideas, which is a crucial component of scientific endeavours,  whose decline we noted in Section~2.)

%==========================================================================
\subsection{What would ``failure'' of research as a social process mean?}
\label{sec:failuretaxonomy}
%==========================================================================

Before we pursue the mechanical analogy, we have to answer a fundamental question about the meaning of ``collapse of science''. Its functions are more complex than those of a bridge, so would we consider science still working if only some of its components are compromised?
  
The relevant functional and operational failures we consider here are all
\emph{creeping}: slow, sub-surface, precursor-laden, and explicitly \emph{not}
the catastrophic exogenous shocks (war, collapse, planetary disaster) that the
acute-crisis literature already covers. Organising by \emph{which subsystem loses
	capacity} yields six such \textbf{failure mode} families (no ordering in terms of importance is assumed).
\begin{itemize}
	\item \textbf{Discovery Capacity Loss (epistemic yield) [DCL].} Breakthrough slowdown beyond
	what logistic saturation of the growth curve would predict; canonical stagnation
	in large fields \citep{Chu2021SlowedCanonicalProgress,Park2023PapersPatentsAre}; and the rising \emph{burden of
		knowledge}, the ever-longer training needed to reach a receding frontier, which
	erodes the capacity to keep up with the most innovative research, as well as the burden of information overload \citep{Jones2004InformationOverloadMessage}.
	\item \textbf{Human Capital and workforce Loss [HCL].} Erosion of career attractiveness and
	social prestige diverts talent away from research; field-level expertise may become extinct due to a
	training hiatus (as in the case of nuclear engineering during long construction pauses in 1980-2010);
	upstream losses from weak STEM schooling and declining higher-education quality;
	loss of \emph{tacit, craft} skill never captured in papers and lost when a
	generation retires \citep{Collins1974TeaSetTacit,Massingham2018MeasuringImpactKnowledge}; and precariat lock-in as insecurity of employment 	becomes structural \citep{Milojevic2018ChangingDemographicsScientific,Spina2022BackZeroPrecarious,OECD2021ReducingPrecarityAcademic,Burton2022AcademicPrecariatUnderstanding}.
	\item \textbf{Knowledge-Base Integrity compromise [KBI].} Failure of self-correction, as
	unreplicated or false findings accumulate faster than they are retracted, aggravated
	because the corrective mechanism (replication) is itself starved of funding, together with metric-gaming pollution of the literature
	\citep{Fire2019OverOptimizationAcademic,Smaldino2016NaturalSelectionBad}.
	\item \textbf{Funding and Resource base Decline inhibiting (parts of) research [FRD].} Disciplinary defunding under political or
	return-on-investment pressure; winner-take-all concentration starving the base;
	and commercial capture that steers agendas toward short-horizon questions (requiring research to have ``impact''.)
	\item \textbf{Social Licence and Autonomy loss [SLA].} Social trust erosion, especially connected to actual behaviours (as contrasted to verbal generalized support); creeping managerial or political constraints or pressures on academic freedom; and the
	politicisation of credibility, where expert status is assigned by tribal loyalty
	\citep{Kahan2011CulturalCognitionScientific}.
	\item \textbf{Internal Control Systems [ICS] failure}. Loss of functionality of peer review, interest in replication, corrections \& retractions.  An epistemic conservatism ratchet, as
	mediocrity-bias selection compounds across cohorts 
	\citep{Boudreau2016LookingLookingKnowledgea,Nicholson2012ResearchGrantsConform,Sobkowicz2015InnovationSuppressionClique}; at the same time the push for ``novelty'' as the differentiating factor; the  overload of peer-review process as submissions outpace reviewers
	\citep{Kovanis2016GlobalBurdenJournal}; and attention/information overload degrading the field's
	ability to track its own output.
	\item \textbf{Consequences of Research-Related Harm to society [RRH].} Growth of the use of technologies and policies based on insufficiently tested research (e.g. in medicine); development and misuse of dual-purpose research and related technologies; diminishing ethical constraints on such research due to funding pressures.
\end{itemize}

\begin{figure}[h]
	\begin{center}
	\includegraphics[width=0.5\textwidth]{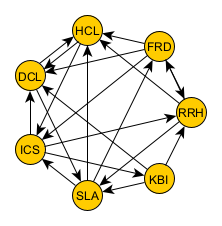}
	\end{center}
	\caption{Positive feedback influences between various failure modes. For explanations of the acronyms, see text. \label{fig:failuremodesnet}}
\end{figure}

The examples of systemic failure may initially occur separately, but they are connected via multiple feedback loops (Figure~\ref{fig:failuremodesnet}). 
Some of the pathways of these feedback mechanisms are described below.
\begin{itemize}
	\item DCL (Discovery Capacity Loss) increases
	\begin{itemize}
		\item HCL, via decreased attractiveness, prestige and interest in research career for top performers.
		\item SLA, via lack of progress leading to perception of wasting money on unnecessary activities/research.
	\end{itemize}
	\item HCL (Human Capital Loss) increases
	\begin{itemize}
		\item DCL, via decreases in the pool of best performers.
		\item ICS, via focus on measurable metrics needed in hypercompetitive environment, and downgrading the importance of ICS.
	\end{itemize}	
	\item FRD (Funds and Resources Decline) impacts
		\begin{itemize}
		\item HCL, via increased competition, precariat,  pressure of conformity within the measured metrics, rather than cultivation of creativity and curiosity.
		\item DCL, via lack of funding for most submitted projects, mediocrity bias in grant schemes, question narrowing, methodological monoculture.
		\item ICS, via decreased funding for activities deemed secondary (replication studies), or push to provide control services (e.g. peer review) for free, and creation of ``low value'' mental association.
		\item RRH, when researchers starved for funds decide to grasp unethical or unsafe research directions; when hypercompetition and the drive for novelty push for the release of untested results (e.g. in medicine)
		\end{itemize}
	\item RRH (Research-Related societal Harm) impacts
	\begin{itemize}
		\item SLA, via examples of harmful research; calls for defunding; calls for decrease of autonomy and imposing external control to avoid such cases.
		\item HCL, via negative associations, loss of social acceptance and prestige of research jobs due to examples of misuse or harm.
		\item FRD, via knee-jerk defunding reactions to individual cases expanding to whole areas of research.
	\end{itemize}	
	\item KBI (Knowledge Base Integrity loss) influences
	\begin{itemize}
		\item SLA, via publicised examples of unverified, false, fraudulent research (when detected).
		\item DCL, by corrupting the starting point knowledge; by enforcing unnecessary repetitive research and vice versa, by discouraging potentially fruitful, but under-recognised directions.
		\item RRH, via loss of (or hiding) information about potentially harmful effects.
	\end{itemize}		
	\item SLA (Social Licence and Autonomy loss) impacts
	\begin{itemize}
		\item ICS, via lesser belief in the capacity of internal control mechanisms and reliance on top-down, state or corporate systems.
		\item HCL, via decreasing chances of long term employment allowing curiosity driven research.
		\item FRD, via imposed decisions on amount of funding and funded research areas, decoupled from academic interests.
	\end{itemize}		
	\item ICS (Internal Control Systems failures) influence
	\begin{itemize}
		\item RRH, directly, via loss of control functions monitoring quality, safety and ethics of the research.
		\item DCL, via allowing misuses/abuses of the system; allowing proliferation of high-volume/low-quality research.
		\item SLA, via decreasing societal trust in the internal, self-correcting mechanisms of science.
	\end{itemize}	
\end{itemize}

The direct, adversarial attacks on science integrity and value may occur in many of the failure modes. 
The easiest target is the social licence and autonomy (SLA), where rumours and accusations are an obvious tactic and calls for external oversight abound. But other modes are susceptible as well: calls for defunding of science (or particular research) impact FRD; portraying scientists as corporate/governmental lackeys impacts human capital; exaggerating social harm (or inventing it) can raise panics (recall the stories of a black hole devouring the Earth when Large Hadron Collider was being powered up, or, more seriously, the successes of  antivaccination campaigns). These attacks are often more visible and acute than the slow, creeping stresses occurring without malevolent actors. This makes them easier to recognise, but not necessarily to counter.

We note that in many cases the growth of harm-leading (RRH) research activities would be attributed to ``locally legitimate'' actors: governments or military agencies pursuing their own goals, formally approved in democracies, but often kept out of ethical monitoring  and conducted without in-depth analyses of potential social consequences. 

The influences described above are, in most cases, nonlinear and non-instantaneous. Many have long recovery times (the effects of a particular failure modes may persist well after a particular impulse has passed.) We shall now turn from the general, often unmeasurable categories, to specific observables that could signal the changes.

But before we do this, we need to mention that not all ``activities'' that form the influence network are necessarily negative. One can imagine specific interventions, strengthening rather than weakening the mechanisms of good research. Increasing funding, so that more projects are pursued. Dedicated journals or funding initiatives that support replications and cross-checking of published results. Changes diminishing ``post-doc slavery'' and precariat of early careers. More acceptance of  curiosity driven  and high risk research. More focus on cleaning the knowledge base, and on publication retractions and their understanding. Coupling internal and external control mechanisms to increase both societal and internal trust and to prohibit harmful research behaviours. There are many such options. In the picture drawn above, in many cases their effects propagate through feedback mechanisms. Unfortunately, we should remember the asymmetry between destruction and repair: the difficulty of the repair task is often much greater than the difficulty of destroying something, and the times of recovery may be significantly longer.

An interesting link connects DCL, SLA and funding. One may ask legitimately: if the individual publication rate is flat \citep{Fanelli2016Researchers’IndividualPublicationa} and the inputs required per genuine advance are rising exponentially \citep{Bloom2020AreIdeasGettinga}, why do societies continue to underwrite an ever-larger research workforce whose marginal breakthrough is ever more expensive and ever harder to attribute?  Is it because no-one knows who/when will make a breakthrough discovery?
Or because the research community has convinced general societies that ``you need science'', just like society funds arts? 
Breakthroughs form a small fraction of research results. By definition, revolutionary  works are unpredictable and the winners cannot be identified in advance, which strengthens the argument for large and broad funding, just to increase the probability of success \citep{Stephan2015HowEconomicsShapes}. However, while the occurrence of high impact work is  difficult to predict, the historical distribution of successful teams shows specific patterns and is quite stable. 
Some institutions are consistently better than the average. Thus there are voices calling for further concentration of funding (even above currently observed one), focusing it on places where the ``real science'' is done.

At the same time, one should not forget that much of what the ``less productive'' researchers and institutions do is not limited to writing papers (not to mention the breakthrough ones), but the covers training people, maintenance of instruments, and provision of standing expertise that constitute a society's \emph{absorptive capacity}, defined as  its ability to recognise, verify, and apply knowledge generated anywhere \citep{Cohen1990AbsorptiveCapacityNew}. A country without an active research base cannot even consume the global knowledge progress, let alone add to it. 

As for the second, more self-serving explanation, it is also quite real. As noted before, the research community, universities, and funding agencies form a coalition aligned via joint incentives, and act accordingly (everyone reports success; no one reports the base rate of failure). In a system where every grant ``succeeds'' and every field insists it is indispensable, such messaging has a strong chance of persuading society of its own necessity. Until the belief breaks down. 

The problem with these two competing explanations is that they are impossible to distinguish, when we measure the results alone. And directly measuring the \textit{intentions} is impossible. A system whose value is diffuse (consists of a mixture of mundane majority and few breakthroughs), and spans long  time horizons is observationally identical to one that has simply used its historical position to continue its funding and persuaded the public of its own necessity: both report success, both insist they are indispensable, and neither exposes a measurable base rate of failure, for fear of turning public opinion against it. 
One could argue that there is a clear distinction between the two mechanisms: the first relies on actual progress (technical, medical, quality of living) attributable to science; the second has no such foundation. But we know that quite universally, the progress is absorbed into the baseline evaluation and almost immediately ceases to be evidence of anything. The achieved benefits become normalised into the baseline (who remembers the time when there were no mobile phones?). Advances in technology quickly stop functioning as proofs of the value of science. This is well known in psychology: we quickly adapt to better conditions. In contrast, losses and threats dominate our attention, while steady benefits fade. There is a second mechanism that discounts the importance of factual proofs of science's value: we are also very good at  motivated reasoning/motivated ignorance \citep{Kahan2016PoliticallyMotivatedReasoningb,Kahan2013IdeologyMotivatedReasoning,Kahan2013IdeologyMotivatedReasoninga,Bolsen2025MotivatedReasoning,Epley2016MechanicsMotivatedReasoning,Pennycook2018LazyNotBiased}. Even a measurable, attributable benefit of science would fail to count if attending to it would force the observer to contradict something relevant to their own identity.

Such a situation is fragile, because exposure of facts contradicting these narratives (or even misinformation to that effect) can dramatically diminish the trust, and with it, funding levels, and the stability of the whole system.

So, to summarise: is there a \textbf{universal answer to the question of what ``collapse of science'' would mean}? We think not. The answer depends on how you \textbf{prioritise} research goals and functions. 
There are multiple possible definitions and scenarios of large scale failure.
It would be different if you value science as part of a system allowing people to pursue their interests, ``curiosity-driven research,'' on a footing similar to the arts: then collapses of funding and social acceptance, and loss of human capital, would be crucial. If you think the role of science is to prepare the ground for new technologies or medical solutions, then the drop in discovery capacity is the main concern. Or if you are worried about potential misuse of scientific results, endangering our freedoms or even lives, then the failure of control systems, and in particular loss of control over potentially harmful research, can be considered a catastrophe. Research would be preserved, but whom it would serve? All of these viewpoints are valid concerns. So, perhaps, instead of looking at science as a whole, we should focus on specific failure modes.

%==============================================================================
\section{Understanding resilience -- analogies from mechanical engineering and ecology}
%==============================================================================

Our approach is based on  understanding  of resilience in complex systems.
We shall be borrowing concepts from two domains where resilience has been studied and modelled extensively: mechanical engineering and ecology. 
Here we are interested in resilience to slowly changing conditions, rather than to acute shocks 
\citep{Boin2008PoliticsCrisisManagement,Bonanno2004LossTraumaHuman,Bruneau2003FrameworkQuantitativelyAssess}.  Such abrupt changes can, of course, influence  research organisations and reorganize the whole process. Drastic examples are provided by changes of functioning and focus of research during wars. 
We should remember, however,  that even the slowly happening changes can lead to abrupt, extreme, even catastrophic events \citep{Albeverio2006ExtremeEventsNature}.

Complex systems can possess
\emph{alternative stable states} separated by tipping points, at which a slow
change in conditions triggers an abrupt, hard-to-reverse regime shift
\citep{Scheffer2012AnticipatingCriticalTransitionsa,Scheffer2009EarlyWarningSignalsa}. Recent work carries this from metaphor to measurement for
\emph{social} organisations: \citet{Schweitzer2022ModelingSocialResilience} develop an agent-based and
network framework that quantifies resilience as the combination of two distinct
capacities, \emph{robustness} (resistance to shocks) and \emph{adaptivity} (the
capacity to reorganise), and monitors it from longitudinal data rather than
judging it only after the fact.

%==============================================================================
\subsection{Mechanical analogy}
%==============================================================================
Imagine a construction (for example a bridge), working under normal conditions. The load does not surpass the designed construction limits. To casual observers (for example the bridge users) everything seems to be  all right. Yet even this normal load may cause slow growth of defects, almost invisible on the surface, which do not immediately affect the operations. When enough of these defects accumulate, two things happen. First, the defect \textit{growth} becomes faster. Second, the defects weaken the structure. If not checked and countered, the tiny cracks, the slow corrosion may eventually lead to the bridge collapsing suddenly, under a perfectly normal load. Examples abound throughout the world.

It is therefore not surprising that the study of creeping disasters has a long history in materials engineering. The ability to predict when an apparently healthy structure may collapse is of crucial importance. 
Of course, owing to the huge range of properties of constituent materials and the details of how they are combined in specific constructions or machines, the discipline is now heavily specialised.
Here, we will focus on a generalized, simplified (yet still complex) approach called \textit{continuum damage mechanics} (CDM). It will provide us with a basic framework (concepts, variables, mechanisms, equations) that, we hope, can be meaningfully transferred to the study of resilience of research.

Materials fail under loads \emph{below} their nominal strength through three slow
modes. \textbf{Creep} is progressive strain under a constant sub-critical load,
passing through primary (decelerating), secondary (steady), and tertiary
(accelerating) stages before rupture. 
\textbf{Fatigue} is the damage accumulated in repeated cycles of stress. One then looks for the relation between the stress amplitude and the number of cycles at which the system fails (the so called N-S curve). 
\textbf{Stress-corrosion cracking} is the acceleration of either of these by an aggressive chemical \emph{environment}. 

Consider first a stationary load, measured by \emph{stress} $\sigma$. 
\emph{Creep strain} $\epsilon$ is
the (initially recoverable) deformation it produces.
Recoverable means that once the load vanishes, the strain (deformation) eventually returns to zero or to some residual value \citep[p. 63-64]{Rosato2001PlasticsDesignHandbook}.

\emph{Damage} $\omega\in[0,1]$ is the irreversible, monotonically increasing loss
of intact load-bearing cross-section, from $\omega=0$ (pristine) to $\omega=1$
(rupture). Its central consequence, due to the accumulation of defects, is that the material feels an amplified \emph{effective stress} $\tilde{\sigma}$ \citep{Rabotnov1969CreepRupture,Kachanov1999RuptureTimeCreep,Krajcinovic1989DamageMechanics}: 
\begin{equation}
	\tilde{\sigma} = \frac{\sigma}{1-\omega}.
	\label{eq:effstress}
\end{equation}
This means that accumulated damage lowers the capacity to withstand stress, or, in other words, increases the effective stress. Above a certain damage level, even ``normal'' stress $\sigma$, below the construction's design limits, turns into $\tilde{\sigma}$ greater than these limits, causing collapse.

Furthermore, the damage itself grows under that effective stress, $\dot{\omega} = f (\tilde{\sigma}, \omega)$.  In the coupled Kachanov--Rabotnov simplified 
approach, both creep strain ($\epsilon$) and damage ($\omega$) evolve together:
\begin{eqnarray}
	\dot{\omega} & = & C \tilde{\sigma}^{\nu} = C\left(\frac{\sigma}{1-\omega}\right)^{\!\nu},\\
	\dot{\epsilon} & = & A \tilde{\sigma}^{\mu} = A\left(\frac{\sigma}{1-\omega}\right)^{\!\mu},
	\label{eq:kr}
\end{eqnarray}
where dot  denotes time derivative.

Equations~\ref{eq:effstress} and \ref{eq:kr} drive the system evolution: as damage accumulates ($\omega\to1$), the
effective stress in Eq.~\ref{eq:effstress} diverges, $\dot\omega$ runs away, and
a slow secondary creep tips into fast tertiary creep and rupture at a finite
time~$t_R$, which can be calculated knowing system parameters.

In an aggressive environment (such as the presence of corrosion or radiation, which pose separate
challenges and complexities at the atomic scale) the damage progresses even in the absence of stress.  Therefore, it enters the macroscopic picture of 
CDM in two ways. First, as a multiplier increasing the 
damage-rate coefficient $C$: corrosion accelerates cracking at unchanged stress. Second, through the damage growth due to corrosion in the absence of stress, which may be described as an additive term  $\gamma \geq 0$ in the equation defining $\dot{\omega}$. 
This is an approximation, because in reality the damage due to constant load can also accelerate the corrosion rate, making the two processes grow in a feedback loop. 

\subsubsection{Multiple damage channels}
There are frequent cases in which damage may accumulate under stress in multiple ways (damage channels).
Two general approaches are often used in such situations.

In metals, an additive approximation postulates that the aggregate damage is the sum of $k=1\ldots N$ 
channel damages, all coupled through a shared effective load:
\begin{eqnarray}
	\omega_{TOT} & = & \sum_{k=1}^{N} \omega_k, \label{eq:addD} \\[2pt]
	\tilde{\sigma} & = & \left(\frac{\sigma}{1-\omega_{TOT}}\right), \\[2pt]
	\dot{\omega}_k & = & C_k\tilde{\sigma}^{\nu_k} +\gamma_k. \label{eq:addrate}
\end{eqnarray}

Every channel sees the same denominator
$1-\omega_{TOT} = 1-\sum_k \omega_k$, so each mechanism accelerates \emph{all} the others through the
common effective load. Overload and corrosion cross--amplify even when their
kinetics are otherwise independent. 

In brittle materials a multiplicative approach may be more useful, strengthening the role of the weakest part (one in which damage grows the fastest). One can introduce here the ``survival capacity'' to withstand stress, defined as 
\begin{equation}
	\Phi =  \prod_{k=1}^{N}(1-\omega_k).
	\label{eq:survmult}
\end{equation}
The effective damage is then given by 
\begin{equation}
	\omega_{EFF} = 1 - \Phi = 1-\prod_{k=1}^{N}(1-\omega_k).
	\label{eq:effdammult}
\end{equation}
The effective stress is then 
\begin{equation}
	\tilde{\sigma} = {\sigma}/{\prod_{k=1}^{N}(1-\omega_k))},
	\label{eq:effdstressm}
\end{equation}
which then determines the damage growth $\dot{\omega}$ via Eq.~\ref{eq:addrate}. We note that for small damages ($\omega_k \ll 1$) both models reduce to the additive case, but they diverge for larger values of damage components.

\subsubsection{Coupling of multiple channels, network approach}
The bridge collapse example used before, while simple and evocative, is not the best one to describe the problems faced by modern day science. One needs a more complex analogy, with multiple components interacting in differentiated ways. Consider then a combustion engine. To properly function, the cylinder and piston must be (sufficiently) mechanically sound, to withstand high pressures and temperatures. The lubrication must be adequate (again, despite high temperatures and particulate residue contamination). Fuel and air must be delivered in the right combination (varied according to the required engine power). 
All these components must function well enough for the engine to do its work. And they depend on each other: deformation of the piston may damage the cylinder, defects in piston/cylinder fitting cause loss of lubricant, wrong fuel/air mixture damages the mechanical part, and finally, too little fuel causes the whole engine to stop.

It is possible to extend CDM by including interactions between the damage channels in the mathematical framework described above. In the simplest version, a given damage channel $j$ may \emph{accelerate} an already-running damage process  $k$, so its time evolution is given by:
\begin{equation}
	\dot{\omega}_k = C_k \tilde{\sigma}^{\nu_k} \Big[\,1+\sum_{j\neq k}\kappa_{kj}\,w(\omega_j)\Big].
	\label{eq:omeganetw}
\end{equation}	
In addition to the intrinsic evolution of $\omega_k$ we have here the summed influence of other damage channels.  $\kappa_{kj}\ge 0$ is the gain from channel $j$ onto channel $k$, and $w(\omega_j)$ an
increasing coupling response with $w(0)=0$. A reinforcing loop
between the channels $k$ and $j$ is positive: both $\kappa_{kj}$ \emph{and} $\kappa_{jk}$ are non-negative (but not necessarily symmetric); and what governs the loop is the product
$\kappa_{kj}\kappa_{jk}$, not either term alone.
The values of $\kappa_{kj} w(\omega_j)$ can be understood as the weights of a directed \textit{network} of damage channels. We shall find a similar, network-based viewpoint in studies of ecological change and collapse. In fact, the network paradigm is a crucial one for modern analyses of resilience \citep{Gao2016UniversalResiliencePatterns,Liu2024AdvancedProgressNetwork}, and fits the complexity of science best. 

\subsubsection{Directed (adversarial) damage}
The mathematical formulation presented above allows to include focused damage effects due to the adversarial activities (found in mechanical engineering, too: someone damaging a specific part of the bridge or engine, affecting at relatively low effort, the performance of the whole structure). 

The best suited for such cases is the networked, multiplicative approach (given by combining equations~\ref{eq:addrate}-\ref{eq:effdstressm} and \ref{eq:omeganetw}). The ``attack'' may be included in the model via 
\begin{itemize}
	\item increase of specific damage component channel $\omega_k$, with catastrophic effects due to the multiplicative definition of effective damage $\omega_{TOT}$. 
	\item increase of channel specific corrosion terms ($\gamma_k, C_k$)
	\item increase of specific network weights $\kappa_{kj} w(\omega_j)$ which could result in cascade spillover of damage from one channel to other channels. 
\end{itemize} 
This flexibility allows modelling of quite complex situations.

\subsubsection{Measuring the subsurface damage}

Lastly, we note that to be able to effectively predict the potential collapse or failure, the engineers have to monitor the structure for hidden damage. Multiple experimental ways are in use to study the progress of the degradation (ultrasonic testing, X-rays, acoustic emission, infrared tomography, eddy current testing, and many more). One of the most widely used methods is \ldots visual inspection, looking for cracks and defects appearing on the surface, no longer hidden beneath. Which seems easy, but one has to know what to look for. 
On top of these measurements, detailed models predicting the time evolution of the resilience have been developed. All these methods are used routinely, and required by law in many countries for crucial constructions or machinery.
One may ask then, why are unpredicted construction collapses still possible? 
At least part of the answer lies in the fact that monitoring subsurface damage progress is cumbersome, sometimes costly, and that many most vulnerable elements of constructions are hard to access. Constant monitoring is required for our safety, a lesson that may be applicable to the health status of science. 

%==============================================================================
\subsection{Ecology analogy}
%==============================================================================

The previous section proposed the damage--accumulation
approach taken from continuum damage mechanics as one way to structure the description of gradual growth of problems in science. This section develops a second,
independent source of analogies and models: the ecology of environmental regime shifts and resilience.
We argue that it fits the creeping crisis of science in the places where the
mechanical analogy falls short. The CDM nicely describes an irreversible ratchet toward
rupture via damage accumulation. In contrast, ecology offers \emph{recoverable multistability}, by considering a system that may occupy one of the plural
alternative stable states, and which can be pushed between them. 
As such, it can in principle return to the original state,
often not directly, but along a hysteretic path. The two approaches are complementary, and their
gaps are informative: ecology lacks a way to include a planning, optimising adversary
that the CDM analogy can provide,
while CDM lacks the basin geometry, hysteresis, and slow--variable structure that
the ecological model supplies. 
Where damage mechanics supplies the variables and their kinetics, ecology supplies
the apparatus for \emph{anticipating} the shift from one state to another. 

Resilience in its foundational
sense is the capacity to absorb disturbance and retain function has been one of foundational notions in ecology 
\citep{Holling1973ResilienceStabilityEcological,Folke2006ResilienceEmergencePerspective,Schweitzer2026ResilienceUnderstandBreakdown}. More generally, it is widely applied to complex systems near a tipping point. In such situations, they exhibit universal characteristics in the form of 
\emph{early-warning signals, EWS}: critical slowing down, rising
autocorrelation and variance, and flickering. 
These are observed across ecosystems,
markets, and in climate studies \citep{Scheffer2012AnticipatingCriticalTransitionsa,Scheffer2009EarlyWarningSignalsa}. 
One of the tasks we foresee in our research directions is identifying measures that could serve as such EWS in the case of monitoring the health of research. 
We have to be careful, though, that  early-warning indicators carry high false-alarm rates and often fail on real data \citep{Dakos2012MethodsDetectingEarly,Dakos2015ResilienceIndicatorsProspects}.
Moreover, the simple fact that a system is resilient should not be confused with positive ``valuation'' of the state. A degraded equilibrium can be highly resilient, so the objective is \textbf{resilience of
\emph{desirable} functions}. Together with the
reflexivity of social measurement (observing the system changes it), these are the principal challenges our research programme must be aware of.

Society is explicitly within the ecological framework's intended scope, 
and the transition between its quasi-equilibrium states needs not be negative. Paths to \emph{positive} social tipping (one leading to a more desirable social system state) are also possible \citep{Otto2020SocialTippingDynamicsa,Walker2004ResilienceAdaptabilityTransformability}.

In the simplest case the mathematical formulation of the ecological framework is via a potential landscape $U$, shaped by a slow driver \citep{Holling1996EngineeringResilienceVersus,Walker2004ResilienceAdaptabilityTransformability}. 
System equilibria are the stationary points of $U$, stable states are its minima. Ecological resilience depends on the geometry of the occupied basin and on how the potential function changes.

\begin{figure}[t]
	\centering
	\includegraphics[width=0.95\textwidth]{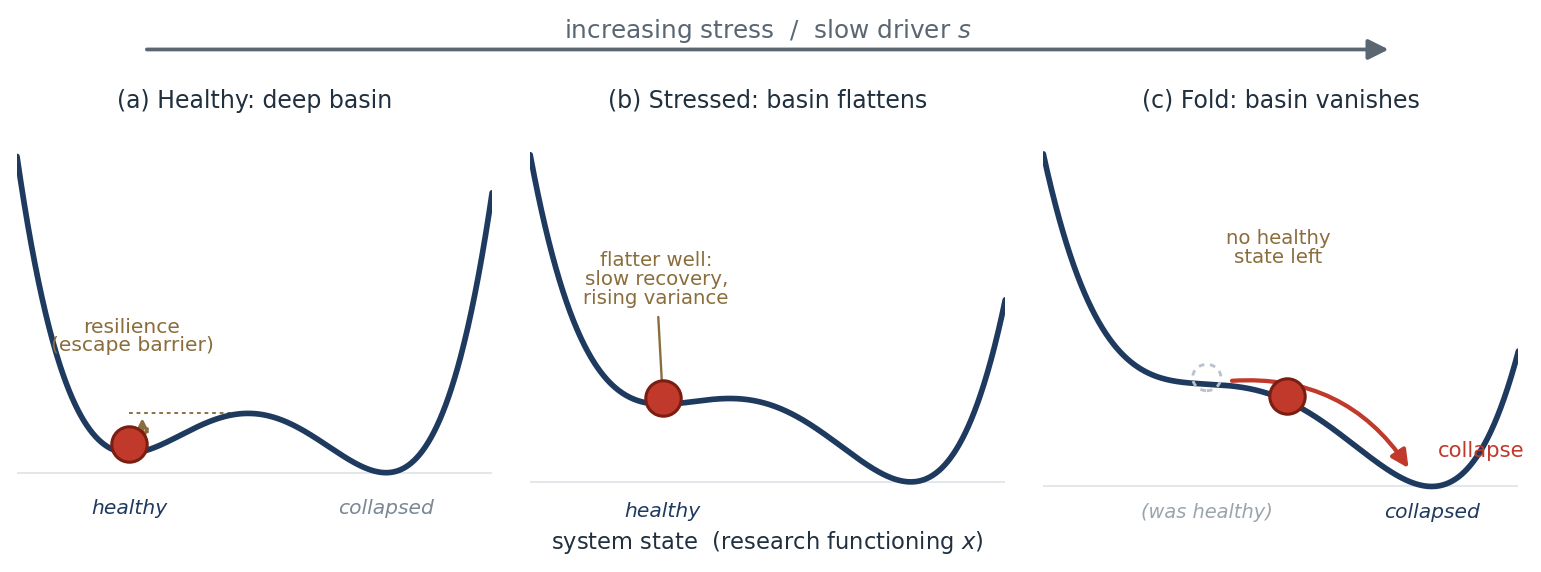}
	\caption{The potential-landscape view of creeping disruption. As a slow driver
		$s$ (accumulated internal and external stress) rises, the basin of the healthy
		state (a) flattens, so recovery from perturbations slows and fluctuations grow,
		the generic early-warning signature (b), until the basin merges with the barrier
		in a \emph{fold} and vanishes (c), and research functioning drops irreversibly to
		a degraded state. The barrier height in (a) is the system's resilience; its slow
		erosion, not any acute shock, is what precipitates collapse. Recovery from the
		collapsed state requires reversing $s$ well past the tipping point (hysteresis).}
	\label{fig:fold}
\end{figure}

The system's behaviour at an equilibrium point is dynamical: it allows fluctuations of its state, as long as they are below the resilience barrier (see Figure~\ref{fig:fold}). In the presence of disturbance, the barrier may be lowered, and eventually the system may collapse to another state. 
As with mechanical stresses, the disturbance may have different time scales and forms. \citet{Lake2000DisturbancePatchinessDiversity} identified three basic forms of the disturbances: 
\begin{itemize}
	\item Pulse — a short, sharp perturbation with a distinct end (a flood, a frost). The system is hit and then left to recover. This is the acute shock we explicitly set aside.
	\item Press — a disturbance that arrives and then persists at a new sustained level (a permanent change in flow, chronic pollution). The stressor doesn't relent; the system settles to a new regime under continuous load.
	\item Ramp — a disturbance whose strength steadily increases over time without a plateau (progressive sedimentation, a slowly worsening drought). This is the form that maps directly onto creeping disruption: a slow driver that keeps rising, exactly the $s$ in the fold figure (Fig.~\ref{fig:fold}).
\end{itemize}
It is worth noting that despite their differences, all three forms of disturbance may lead to abrupt system changes. What changes is often the lag time between the pulse or onset of press/ramp and the system collapse.
In our ``creeping failure'' approach  the press and ramp stressors are most relevant.

For our analysis, the dynamics are best described through dividing the system into fast and slow states, $f$ and $s$  \cite{Rinaldi2000GeometricAnalysisEcological,Rocha2018CascadingRegimeShifts}. General formulation of the evolution equations may be then written as 
\begin{eqnarray}
	\dot{f} & = & F(f,s)\\
	\dot{s} & = & \varepsilon S(f,s),
	\label{eq:M1}
\end{eqnarray}
where dot denotes time derivative, functions $F$ and $S$ depend on the details of the described system, and $\varepsilon \ll 1$ separates the slow (environmental) changes from the fast ones (e.g. species population).

There are multiple detailed applications of the models to specific systems \citep{Tilman1994HabitatDestructionExtinction,Scheffer2001CatastrophicShiftsEcosystems,Hirzel2008HabitatSuitabilityModelling,Scheffer1990MultiplicityStableStates,Scheffer1989AlternativeStableStates,Scheffer1993AlternativeEquilibriaShallow,May1977ThresholdsBreakpointsEcosystems,Mayer1995IntegrativeModelOrganizational}. 
In most cases what is studied are transitions between different equilibria of the described system.
In the model language, the creeping crisis lives in the gap
between the two resilience basins. As the slow factor $s$ changes, the initially occupied minimum in $U$ flattens and eventually merges with the adjacent saddle in a fold, at which point the basin
vanishes and the state collapses to the alternative, adjacent minimum. 

It is also possible to describe a multi-variable, networked version of equations \ref{eq:M1}. 
Assume that the system is described by sets of fast and slow variables $f_i$ and $s_i$. 
The dynamics of fast variables could then be expanded to
\begin{equation}
	\dot{f_i} = F_i(f_i,s_i) + \sum_j G_{ij}(f_j,s_j),
\end{equation}
so that changes in each state variable may influence other variables.
The resilience and stability of ecological networks was developed to describe the real world, complex ecosystems \citep{Sole2001ComplexityFragilityEcological,Memmott2004TolerancePollinationNetworks,Guimaraes2020StructureEcologicalNetworks,Lenton2008TippingElementsEarth’s}. The meaning of the fast/slow variables $f_i, s_i$ and the specific forms of the functions $F_i, S_i, G_{ij}$ depend on the environmental system under study. 
At the same time,  the universality of the approach makes it suitable to study social systems, like the state of science and its resilience.

%==============================================================================
\section{The basic framework of the proposed model}
%==============================================================================
As noted in the opening sections, the signals that may (or may not!) indicate the deterioration of science as a social process are quite varied. Fortunately, both disciplines from which we intend to borrow and modify the existing modelling apparatus already allow networked, multiple damage channels and health measures. We are going to present an initial choice of the dimensions (channels) that might serve as descriptors of the collapse process.

\subsection{Four key measurable dimensions of the collapse}
The six aspects of functioning research, identified as areas of potential failure modes in Section~\ref{sec:failuretaxonomy} are quite general, and internally diverse. Some contain observables that are quantifiable (and relatively accurately measurable). Others, in particular, Knowledge Base Integrity [KBI]  and Research Related Harm to society [RRH] are very difficult to measure, so they are often treated as qualitative. 
To be able to creatively re-use the notions and techniques offered by mechanics and ecology, we must focus on a selected few ``health indicators'' that would cover most of the aspects of potential creeping corruption, and be precise enough to feed the models adapted from the earlier resilience studies.

Following the exploration--exploitation logic of organisational
learning~\citep{March2009ExplorationExploitationOrganizational}, and treating (where possible) capacities as \emph{intensive}, per-capita
densities rather than raw counts, we resolve the state into four coupled
dimensions: discovery, verifiability, social standing/trust, and human resources potential.
Moreover, as we do not know what are the absolute ``optimal values'' of a truly ``healthy state'' (so that we do not know if the values measured today are ``good'' or ``troublesome''), we will be focusing on the evolution of their levels. In mechanical analogy our interest is on $\dot{\omega}_k$ rather  than on $\omega_k$; in the ecology approach, on time derivatives of $f, s$. 

\paragraph{State variables.}
Table~\ref{tab:fourvars} lists four initially identified measures. 
Three are intensive ratios and one is an
absolute sociological measure; keeping their nature explicit matters, because it fixes what the
meanings are and what possible coupling between them exists. Moreover, these measures depend on the research field (and, possibly, the country).
\begin{itemize}
	\item $x_1$: \emph{exploratory capacity density}, breakthrough discoveries per
	researcher (breadth and vitality of open programs). Observed via the CD/disruption
	index, atypical combinations of terms and ideas signifying novelty, and topic diversity.
	\item $x_2$: \emph{verification capacity}, the share of output that is
	replication, error--correction, and cumulative refinement, as a fraction of total
	publications (dimensionless, bounded). Observed via replication success and
	retraction/correction dynamics.
	\item $x_3$: \emph{trust/legitimacy capital}, public and institutional trust,
	autonomy, and norms, as an \emph{absolute} index. Observed via TISP, autonomy
	indices, and funding stability. Trust is a stock held by the surrounding society; it is not a ``trust per researcher'' intensive measure. Moreover, as we noted, using global average measures is misleading. Distrust differs across research disciplines, and certainly across population segments. Using the mechanical analogy, it is a corrosion process that attacks the weakest points of the system (welding joints, cracks created by mechanical stress).
	\item $x_4$: \emph{human--capital density}, attractiveness of research as a career choice for the most suitable candidates and availability of such a career (positions, funding, stability). Some possible measures that can be used: the trained--people pipeline (incoming/outgoing) relative
	to the current number of researchers (a per capita replacement ratio). Observed via
	PhD--to--faculty transitions, precarity, and mobility/emigration.
\end{itemize}
Two out of these four ($x_1, x_2$) are by definition bounded between 0 and 1. In fact, the exploratory and verification capacities are part of the same ``production pool'', competing for the same resources (the missing part, $z=1-x_1-x_2$, is the research production that is neither truly innovative and important, nor actually verifying the previous results). The third variable, compressing various measures of social trust may also be limited to the 0-1 range (we remind here that the trust may significantly vary between research fields). The last variable is again a complex composite of various aspects of researcher career paths that make research an attractive and sustainable option for the most brilliant and creative candidates. The exact way of combining specific observational measures into the four variables shall be determined in the future. We note here that the engine analogy in Section~6 was only illustrative and there is no direct mapping between the engine components and the $x_i$ variables.

\begin{table}[htbp]
	\centering
	\small
	\renewcommand{\arraystretch}{1.25}
	\begin{tabular}{@{}c p{3.5cm} p{2.9cm} p{4.6cm} m{2.0cm} }
		\toprule
		\textbf{Symbol} & \textbf{Dimension} & \textbf{Nature} & \textbf{Observable(s)} & \textbf{Associated failure modes}\\
		\midrule
		$x_1$ & Exploratory capacity & Intensive (per researcher) &
		CD/disruption index, novelty, topic diversity & DCL \\
		$x_2$ & Verification capacity & Fraction of output ($\in[0,1]$) &
		replication success, retraction/correction & KBI\\
		$x_3$ & Trust/legitimacy & Absolute index &
		TISP, autonomy indices, funding stability. Typically measured in some Likert scale, but can be normalised to [0-1] range. & SLA \\
		$x_4$ & Human capital & Intensive, properly scaled combined measure (per researcher) &
		PhD$\to$faculty transitions, rate of precarious employment, forced mobility, churn, psychological job satisfaction measures. & HCL \\
		\bottomrule
	\end{tabular}
	\caption{The four dimensions of research functioning, their nature, and their
		observables. Funding and Resource Base [FRD] and Internal Control Systems [ICS] are not directly measured by the $x_i$ set, but will act as stressors or coupling parameters in the model. }
	\label{tab:fourvars}
\end{table}

These initially chosen dimensions may, of course, be further expanded or modified. The choice, while not perfect, has one advantage: it allows us to sketch the drafts of implementation of analogies of resilience models used in mechanical engineering and ecology.

Out of the four dimensions, the one most vulnerable to adversarial attack is trust and social legitimacy (where most ``attacks'' today are observed). The least exposed is the exploratory capacity. The two remaining dimensions are affected indirectly, mostly via institutional pressures (or lack thereof), like institutional reactions to specific misconduct cases blown beyond measure by instigated public outrage, anti-scientific campaigns affecting researchers, or lack of support for replication/validation studies prompted by calls for ``innovative'' research. 

%==============================================================================
\subsection{Classifying the social signals}
%==============================================================================

%\subsection{Signals examples}
In addition to the proposed four main system variables, there are more 
``warning signals'' (listed in Sections~2-4). 
A part of the proposed research is looking into measurable quantities associated with these signals, that could be used as additional factors driving the dynamics of the degradation process and possible actions to counter/reverse it. Below we provide partial list of such observables and their characteristics. First, however, we have to note that not every ``negative'' observation is a sign of damage: for example a decline in the number of breakthrough papers in a sub-discipline might not be due to decreasing creativity, but simply to growing maturity of the field in question. The case of methodological monoculture may be similar. A well established methodology that provides robust results will ``monopolize'' a discipline.  In each such case we would thus need to separate damage growth threatening failure from normal fluctuations or processes, typical for specific field.

Table~\ref{tab:observables} presents examples of the observables, together with the measurable signals. The \textit{Stressor type} column indicates the origin of the stress being measured (as defined in Sections~2--3): internal stress, external stress and external corrosion. The \textit{Repair} column indicates an estimate of the reversibility of the damage caused by the stressor: whether the recovery process is fast, slow (slower than the damage creation), or essentially impossible. The \textit{Observation lag} column estimates how long it takes for the observable to register the process in question, for example the bibliometric measures have a high lag time, encompassing single or even multiple publication cycles. 
The Evidence tier column denotes how the measure is derived from observable data. \emph{Tier~1} are documented measures that  can be filtered from data
directly. \emph{Tier~2} are  partial
proxies, inferred from the data indirectly and subject to  observation noise and validity caveats. \emph{Tier~3} are  latent, weak or absent proxies (such as the institutional memory, chilling--effect magnitude, tacit--knowledge loss) that are \emph{not} identifiable from their own data but rather constructed, dependent on model assumptions.

A special comment concerns funding. As noted in Section~\ref{sec:funding}, the global per-capita funding is relatively stable. It is, therefore, quite unsuitable as a dynamics measure. Which does not mean that funding is not important, quite the contrary. But to understand its role we must look beyond the averaged values. What counts is the \textbf{distribution of funding}.   

Some disciplines, and some researchers, teams,  and organisations receive more than others. 
The grant-based system is \textit{ designed} to produce such effect: ``the best projects get funded''. But as we discussed, the selection process is far from optimal and objective. There is a fundamental weakness present: grant evaluation deals with \textit{promises}, not results. To look at it from the resilience perspective: the lack of funding stresses some components (research fields and associated communities) more than the others. In mechanical analogy, some parts of the engine are under much higher load than others, and are thus more prone to collapse. 

Complementary argument related to funding concerns the distribution of temporary (grant-based) funding and long-term one, which allows planning, security of employment and continuation of research. The increased role of temporary funds puts stress not only on the human capital $x_4$, but also on the mixture of exploratory and verification capacities ($x_1, x_2$). The pressure to get new grants results of frequent subject hopping by research leaders/teams, and even more frequent dismissal of validation/replication studies as unproductive (when productivity is understood in term of money).

Special attention should be paid to the role of social trust in science as a failure channel. 
Studies of such trust/distrust dynamics differentiate multiple dimensions of trust: the public separately evaluating the competence, integrity, and benevolence of scientists and research institutions (Muenster Epistemic Trustworthiness Inventory, METI, \citep{Hendriks2015MeasuringLaypeople’sTrust}). 
\citet{Cologna2025TrustScientistsTheir} uses the TISP `Trust in Science and Science-Related Populism' project scale, which adds openness/transparency dimension. It is thus tempting to separate the trust decline signal into these dimensions. However, at this stage we refrain from this, due to the fact that it is quite difficult to separate specific dimensions in real cases. For example the rise in public distrust in response to information about scientists' lack of attention, dishonesty, or malpractice (all very different) is hard to attribute to a single dimension.

\begin{table}[htbp]
	\centering
	\footnotesize
	\renewcommand{\arraystretch}{1.25}
	\begin{tabular}{@{}p{2.8cm} p{1.5cm} p{1.8cm} p{7.0cm} p{1cm}}
		\toprule
		\textbf{Observable, associated failure mode}  & \textbf{Repair time} & \textbf{Observation lag} &
		\textbf{Examples of observable signal(s) or measurement approaches} & \textbf{Evidence tier} \\
		\midrule
		\multicolumn{5}{@{}l}{\emph{Internal stress}}\\
		\addlinespace[2pt]
		Question narrowing [DCL]&  slow & medium & decreases in topic diversity or entropy,
		CD/disruption index, atypical--combination novelty & 1 \\
		Methodological monoculture  [DCL]& slow & medium & increase of dominant method \& term share and loss of diversity in published papers & 2 \\
		Integrity erosion / questionable research practices [ICS,DCL,KBI] & partial or irreversible & long & increase in retraction rate, replication--failure rate, plagiarism rates, image--duplication and paper--mill flags[KBI,ICS] & 1--2 \\
		Evaluation gaming  [ICS]& fast & short & coercive and self--citation, citation--cartels and their 
		detection, Impact Factor manipulation, grant system manipulation cases (presence of cliques) & 2 \\
		\addlinespace[3pt]
		\multicolumn{5}{@{}l}{\emph{External stress}}\\
		\addlinespace[2pt]
		Security in funding availability [FRD] & reversible & short & decrease in proposal success
		rates across disciplines, project topic continuation ratio, stability of project research teams, growth of institutional award-ratio Gini & 1 \\
		Inequality in funding directedness [FRD,SLA] & reversible & short & changes in distribution agency portfolio statistics, discipline inequalities in success	rates, individual inequalities in funding, basic/applied ratio, award--topic Gini & 1 \\
		Political agenda impact [FRD,SLA]& variable & short to medium & increases of mass grant terminations, 		banning of topics, terms in funding directives, agency guidance changes & mix \\
		Human capital loss [HCL]& slow  or irreversible & medium & decreasing tenured/untenured ratios, increase in forced mobility of junior researchers, and researcher dropout numbers & 2 \\
		Administrative load [HCL]& reversible & med & increase on ``wasted'' time, measured by faculty time--use surveys, growth of admin-to-research staff ratios & 2 \\
		\addlinespace[3pt]
		\multicolumn{5}{@{}l}{\emph{Corrosion (load--independent)}}\\
		\addlinespace[2pt]
		Societal trust [SLA] & slow & low & directly via TISP/METI measurements; indirectly via monitoring of social media, and, in selected cases of the prevalence of actual decisions defying scientific consensus & 1 \\
		Norm erosion  [ICS,HCL,SLA]& slow & long & decreased adherence of norms, as measured by norm/counter--norm attitude surveys, increase of retractions due to intentional norm-breaking activities (e.g. image manipulation, data falsification) & 2 \\
		Institutional memory loss [KBI]& irreversible & very long & specific case in-depth analyses of effects of infrastructure decommissioning,
		long--term dataset/cohort loss, mentorship--chain gaps & 3 \\
		Epistemic pollution  [KBI,ICS]& hard & emerging & increases in detected paper--mill examples, fake references rates, LLM-generated text prevalence & 2 \\
		\bottomrule
	\end{tabular}
	\caption{Candidate damage stressor measurement variables for  research as a social process.  
		In the Political agenda case evidence tier ``mix'' denotes documented events but inferential
		magnitude. The trust reversibility may be different for the different METI/TISP dimensions, for example it is asymmetric for integrity \& benevolence dimensions, but reversible for the opennes dimension. Repair times and observation lags are author's estimates.}
	\label{tab:observables}
\end{table}

\subsection{`Localised models': research discipline and inter-country differences}
So far we have treated ``science as a social activity'' as a single system, whose state might be healthy (despite the negative symptoms) or not. The goal was to study the resilience to creeping changes in conditions, both internal and external. The four variables $x_i$ were conceived as calculated or measures on a global scale.

However, there are obvious differences between scientific disciplines/research fields and between countries. The same variables $x_i$, when limited to a specific country and research discipline, may have not only very different values, but also differ in their dynamics. Science can be ``healthier'' in some research fields or countries. Simple global averaging is almost certain to obscure important factors and make the answer to the question ``does science face a potential collapse?'' wrong or meaningless. In addition to the separation of the damage channels, it therefore makes sense to identify specific ``environmental niches'', in which the evolution paths might be quite different.

This division into separate ``ecosystems'' has several important consequences.  First, it is not a strong separation. While the education levels, organisational forms of research, main research interests, funding levels, experience in running large (international) projects may differ between countries, researchers are very mobile (compared with whole populations). This mobility impacts the capacity for research, increasing it in countries that attract the best researchers, decreasing in those that ``lose'' their best (commonly referred to as the brain drain). So, while at some level one can analyse the status and resilience of research in a given country, there are links between them, similar to the networked ecosystems (Eq.\ref{eq:M1}). Also, we have to remember that the research results (publications) face global competition, not just a local one, so local measurements of creativity (variable $x_1$) should be done with respect to global values. At the same time, focusing on a single country has the advantage of much better controlled empirical data: funding, employment, internal mobility.

The division into research disciplines offers similar advantages and challenges. Domains relying on costly infrastructure are different from the ``paper and pencil'' ones. Fields where industrial uptake of results is high (engineering, medicine, computer sciences) differ from fundamental research (high energy physics, astronomy, archaeology\ldots). STEMM research differs from the humanities. Thus many processes and characteristics will be different and allow separate analyses, for example levels of funding devoted to the ``preferred'' and ``secondary'' disciplines. Also, social perception (trust) might be radically different. Breakthrough discoveries rates differ. Similarly, the internationalization levels, publication and citation patterns, and career opportunities.

Still, in most countries, the research is organised into multidisciplinary universities, and in many cases perceived, by the public and governing bodies as a coherent whole (often lumped together with the higher education functions). While adversarial attacks on the position of science are mostly focused on specific disciplines, they impact science in its entirety. Breaches of ethics spill from individuals to  the whole  research system, prompting the calls for deep corrections, without real understanding of the consequences of the proposed reforms. 

%==============================================================================
\section{Conclusion: A research agenda}
%==============================================================================
We now return to the central question: what is the state of health of research as a social activity? Are the undeniable problems listed in the opening section a natural and inevitable part of the social process (like virus or bacteria infections are part of life, or bankruptcies are part of economy), or are they a symptom that in the past decades something went wrong? Institutionalised science, as we know it, is quite young, has achieved so much that the world today is radically different than 350 years ago. Maybe the job insecurity, dropping per capita innovativeness and other observations are only growing pains? Or do they signal a coming collapse?

We cannot answer this question here and now. The available data is too unstructured, too fragmentary. 
But it would be imprudent to disregard the symptoms. The lessons from resilience studies show, that with proper identification of the nature of the disruption, it is possible to mitigate and even reverse the damage accumulation, and avoid collapse.

We suggest therefore a research agenda that would build up and properly structure empirical data, and construct adequate models, allowing us to answer the central question, following the path suggested by \citet{Schweitzer2026ResilienceUnderstandBreakdown}. 
In our opinion the best way to treat science as a (very) complex phenomenon is to account for the  various signals and related subsystems/processes as network of networks. Moreover the networks would likely best described as multi-layer ones, with different types of links between the components. We can then re-use the methods developed in specialized fields, like mechanical engineering or ecology to understand the dynamics in a way that offers some prediction and planning possibilities.

The future work may be composed of:
\begin{enumerate}
\item \textbf{Specify the conditions corresponding to ``the state of health'' of science.} 
Candidate state variables might include the four identified above ($x_i$), but this choice is preliminary, and may be expanded by variables related to: growth rate of the knowledge base, 
actual use of scientific authority in policy formulation, technical/economic usage, 
funding levels and usage, talent 
inflow/outflow, effectiveness of research quality control measures and consensus recognition. 
\item \textbf{Create the foundations of mechanistic and ecological models.} The coupled  system of proposed variables $x_i$ requires defining their \textbf{dynamical evolution equations}. Work on formulation of such driving equations in specific damage channels. Re-analise existing data with such point of view in mind.
\item Rather than attempting to handle the whole complex system immediately, we suggest \textbf{selecting  initial subtopics of the research on science resilience, for example the changes in societal trust in science.} In this case, we define the shift of trust as \textbf{loss of science role as an authoritative input to collective decisions}, and not as ``society rejecting facts'' nor the verbalised support for abstract ``science''. This would involve study specific examples of the loss on a \textbf{per country} and \textbf{per discipline} basis.
\item \textbf{Select ``small project'' candidates}: mechanisms that are ``separable enough'' to allow a dedicated study, so that the proposed research  programme becomes manageable. In this way the crucial data may be gathered and network of interrelationships understood better, but do it in a way that allows integration of the results between the studies into a broader framework. . 
\item \textbf{Search for  candidate early-warning signals (EWS).}   \citep{Dakos2012MethodsDetectingEarly}. Define the measurements and observables  that can be used in this context. Once EWS are found, estimate the level of the danger of collapse.
\item  Develop realistic \textbf{Agent Based Models (ABMs)} of particular modes of failre. A step further would be the development of more detailed ABMs of internal and external processes, along the lines explored in \citet{Sobkowicz2017UtilityImpactFashion,Sobkowicz2015InnovationSuppressionClique}, extended with an explicit adversary \citep{Sobkowicz2021AgentBasedModel}. Import threat modelling and red-teaming from
epistemic security \citep{Seger2020TacklingThreatsInformed} to stress-test the system from the attacker's perspective, the component absent from mechanics and only partial in ecology.
\item \textbf{Study interventions and positive tipping.} Move from diagnosis to design of strategies building resilience. For example, in the case of public trust consider optimal ways to use 
inoculation, institutional trustworthiness, and levers for \emph{positive} social
tipping \citep{Otto2020SocialTippingDynamics}. In the case of productivity think about changing the current careers, motivation and funding systems. Once an operational model is sufficiently accurate and realistic, such policy changes would no longer rely on guesses, but could be called ``evidence driven''.
\end{enumerate}

%==============================================================================
\paragraph{Limitations}
%==============================================================================
The cross-domain transfer from ecology and mechanical engineering is based on analogies: fatigue/creep and ecological EWS are sources of hypotheses and measurements, not laws of social systems. 
Early-warning signals present in these domains may not exist, or may not be detectable in societal systems, for the transitions of interest. They may be obscured by multiple effects of trends and events not directly related to research as a social process: economic crises, wars, pandemics, natural catastrophes, as well as by variances in cultural heritage and socioeconomic systems.

The framework's normative core: defining measures describing which functions of science are worth making
resilient (fundamental research and understanding progress, research that is directly useful to societies, economy driven R\&D programs) may be contested in the light of political priorities and must be argued, not assumed. 

Lastly, the empirical programme our vision implies is very broad and requires multinational collaboration; this paper sets the direction rather than delivering the evidence.

\phantomsection
\addcontentsline{toc}{section}{References}
%\bibliographystyle{unsrtnat}
%%\begingroup
%%\renewcommand{\url}[1]{}%       % swallow the URL text
%%\makeatletter
%%\@ifundefined{urlprefix}{}{\renewcommand{\urlprefix}{}}
%%\makeatother
%\bibliography{../../../../../TeX/Bibliography/Consolidated_25.09.2024.bib}
%%\bibliography{research_resilience_creeping_crisis_draft_V2.bib}
%%\endgroup

\end{document}